# Cobalt Binary Compounds for Advanced Interconnect Materials

Gyungho Maeng[1] and Yeonghun Lee[1,2,3*]

**Abstract**
The industrial standard copper (Cu) interconnects face a substantial resistivity increase at thinner linewidths, posing a well-known challenge to limit overall device performance. To address this issue, we have evaluated the potential properties of cobalt (Co) based binary compounds as replacements for Cu. Co is considered as a promising alternative due to its potential for enhanced reliability and low resistivity at sub-nanoscale dimensions. Furthermore, the combination of elements provides a possibility to engineer novel properties, transcending the limitations of elemental metals and expanding the search space for next-generation interconnects. In this study, a high-throughput screening method was used to identify several Co-based binary compounds with superior electronic transport and reliability at reduced thickness. The findings demonstrate that specific Co-based binary compounds hold significant potential to overcome the performance limitations of scaled interconnects.

## 1 Introduction

Copper (Cu), the industry-standard material adopted through dual-damascene processing, now faces a critical challenge as the metal pitch shrinks below sub-nanometer. The primary issue is a sharp increase in resistivity due to enhanced surface and grain boundary scattering at nanometer scale thicknesses [1–4]. The long electron mean free path (MFP) of 39 nm, which provides Cu its low bulk resistivity, becomes a liability at the nanoscale [5]. Consequently, at a 10 nm line width, Cu exhibits a resistivity that is about an order of magnitude higher than its bulk value [6–8].

Furthermore, copper interconnects require thick barrier layers to prevent diffusion and liner layers for adhesion [9, 10]. However, these layers are non-scalable parameters, which reduce the proportion of the core metal as the linewidth decreases[11, 12]. In contrast, Cobalt (Co) offers the potential for thinner barrier layers due to its higher activation energy for diffusion [13]. Moreover, its relatively short MFP of 19 nm based on experimental results leads to a smaller increase in resistivity at thin thicknesses, making it a promising candidate for next-generation interconnect materials [7, 14].

While the properties of elemental metals are well-established, binary compounds provide a pathway to engineer novel properties by tuning their composition. This vastly expands the materials design space beyond the discrete options offered by pure elements. In the context of interconnect technology, binary compounds offer distinct integration-driven advantages over elemental metals. Prior studies have demonstrated that specific compound phases can exhibit superior adhesion to dielectrics, suppress atomic diffusion, and self-forming barrier or liner behavior [15–20]. These attributes can enhance reliability and enable reduced barrier/liner thickness which is critical at extremely scaled dimensions where interfacial effects dominate the effective resistance. Motivated by these integration advantages, we extend the search for next-generation interconnects to Co-based binary compounds. We specifically focus on ordered stoichiometric phases rather than disordered solid solutions. Since disordered structures experience alloy scattering, they are expected to exhibit lower conductivity than ordered compounds [21, 22]. Therefore, we study ordered compounds as they represent the theoretical conductivity upper bound for a given system; if an ordered compound does not outperform elemental Cu, its disordered alloy counterpart is unlikely to do so. Furthermore, in the context of alloys, the absence of translational symmetry necessitates more advanced modeling approaches, such as special quasirandom structures [23]. Our objective is to systematically explore the expanded material space and evaluate the potential of binary candidates against the industry standards.

We focus on two key figures of merit to address performance and reliability. The first is the product of bulk resistivity and mean free path ($\rho_{\alpha\beta} \times \lambda$), which quantifies intrinsic resistivity scaling behavior. This property is derived from the electronic structure, where a lower value corresponds to more favorable

* Yeonghun Lee
 y.lee@inu.ac.kr
 orcid.org/0000-0002-6058-1316

[1] Department of Electronics Engineering, Incheon National University, 119 Academy-ro, Yeonsu-gu, 22012, Incheon, Republic of Korea
[2] Department of Intelligent Semiconductor Engineering, Incheon National University, 119 Academy-ro, Yeonsu-gu, 22012, Incheon, Republic of Korea
[3] Research Institute for Engineering and Technology, Incheon National University, 119 Academy-ro, Yeonsu-gu, 22012, Incheon, Republic of Korea

conductivity performance at reduced dimension, making it a scalable indicator for superior scaling [7, 24]. The second is cohesive energy ($E_{coh}$). A higher $E_{coh}$ signifies stronger interatomic bonding, which enhances reliability by suppressing electromigration and diffusion [7]. To efficiently navigate the vast compositional space of binary compounds, we employ high-throughput screening based on first-principles calculations to systematically evaluate these two metrics. This study thereby identifies novel binary compounds predicted to overcome the scaling and reliability challenges of future interconnects.

## 2 High throughput screening

Our theoretical investigation was conducted using high-throughput first-principles calculations based on density functional theory. All calculations were performed with the Vienna Ab initio Simulation Package (VASP) [25, 26]. The exchange-correlation functional was described by the Perdew-Burke-Ernzerhof (PBE) formulation within the generalized gradient approximation (GGA) [27, 28]. The pseudopotential is given by the projector-augmented wave (PAW) method [29, 30]. The effects of spin polarization were considered in all calculations.

In order to efficiently explore a wide range of materials, we sourced the initial crystal structures from the Materials Project database via its Application Programming Interface (API) [31, 32]. Candidate compounds were systematically screened based on the following three criteria, as illustrated in Fig. 1: (1) metallic character (band gap $E_G \leq 0$ eV); (2) thermodynamic stability (energy above the convex hull $E_{Hull} \leq 20$ meV/atom); and (3) structural simplicity ($N_{atom} \leq 20$ atoms in the primitive unit cell) [33].

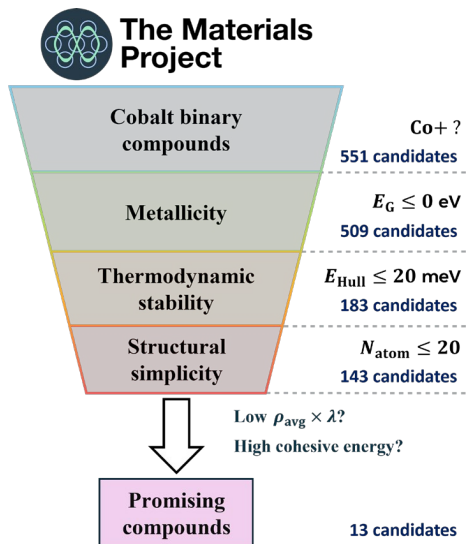

**Fig. 1** High-throughput screening procedure for promising cobalt binary compounds.

For each candidate that met these criteria, a static self-consistent field calculation was performed using standardized VASP input files generated by the Pymatgen library [34]. The initial structures were retrieved from the Materials Project database, where they had been fully relaxed according to their convergence workflow (ionic step energy convergence $\leq 0.0005 \times N_{atom}$ eV, where $N_{atom}$ refers to the number of atoms). To ensure consistency with these pre-relaxed structures, we adopted calculation parameters identical to the Materials Project standards input sets from Pymatgen. Consequently, we proceeded directly with the electronic structure calculations without further relaxation. The energy cutoff for the plane-wave basis set is 520 eV, and k-point density is 100/Å$^{-3}$. We also selectively applied Hubbard U values to $d$- and $f$-orbitals for compounds containing fluorine or oxygen atoms according to Materials Project. To properly incorporate magnetism in Co compounds, the initial magnetic moment on each atomic site was set to the calculated values provided by the Materials Project database; then, the per-site moments were allowed to relax self-consistently. A subsequent non-self-consistent field calculation to analyze transport properties was performed on a significantly dense k-points mesh of 50,000 points per reciprocal atom.

The transport properties were evaluated using the BoltzTraP2 code [35, 36]. The expression for the conductivity tensor is given by:

$$\sigma_{\alpha\beta} = \frac{e^2}{8\pi^3}\sum_n \int_{BZ} d^3k\, \tau_n(\mathbf{k}) v_{\alpha,n}(\mathbf{k}) v_{\beta,n}(\mathbf{k}) \frac{df_{FD}(E)}{dE}\Big|_{E=E_n(\mathbf{k})}, (1)$$

where $\alpha, \beta$ are tensor indices, $e$ is the elementary charge, and $\tau_n(\mathbf{k})$ and $v_{\alpha,n}(\mathbf{k})$ are carrier relaxation time and group velocity for each tensor direction, respectively. The $n$ is a band index, BZ indicates the integration over the Brillouin zone, and the derivative of the Fermi-Dirac distribution function $\frac{df_{FD}(E)}{dE}$ is the Fermi window. The Fermi-Dirac distribution is evaluated at 300 K. Under the constant mean free path approximation, the relaxation time is expressed as $\tau_n(\mathbf{k}) = \lambda/|\mathbf{v}_n(\mathbf{k})|$ where $|\mathbf{v}_{n(\mathbf{k})}|$ is magnitude of group velocity [37]. Substituting this into Eq. (1) yields the conductivity tensor per mean free path:

$$\frac{\sigma_{\alpha\beta}}{\lambda} = \frac{1}{\rho_{\alpha\beta} \times \lambda} = \frac{e^2}{8\pi^3}\sum_n \int_{BZ} d^3k \frac{v_{\alpha,n}(\mathbf{k}) v_{\beta,n}(\mathbf{k})}{|\mathbf{v}_n(\mathbf{k})|} \frac{df_{FD}(E)}{dE}\Big|_{E=E_n(\mathbf{k})}. (2)$$

Based on this relation, the calculated quantity $\sigma_{\alpha\beta}/\lambda$ corresponds to the inverse of the $\rho_{\alpha\beta} \times \lambda$. We use $\rho_{\alpha\beta} \times \lambda$ as the intrinsic figure of merit to evaluate the scaling potential of thin-film resistivity.

## 3 Simulation Results and Discussion

To identify promising candidates for next-generation interconnects, we performed a high-throughput screening of 551 Co-based binary compounds using the first-principles workflow described in Section II. An initial screening based on metallic character, thermodynamic stability, and structural simplicity narrowed the material space to 143 candidates. These candidates were then evaluated based on our two primary figures of merit: $\rho_{avg} \times \lambda$ and $E_{coh}$. The $E_{coh}$ values used in this study were adopted from the Materials Project database. This rigorous process identified 13 compounds predicted to possess superior performance and reliability characteristics.

To represent the resistivity property as a single scalar value,

**Table 1** Most promising Co Binary compounds

| Chemical formula | Material ID | Crystal symmetry | Experimentally Synthesized [a] | Total magnetization ($\mu_B$) | Cohesive energy [b] (eV) | $\rho_{xx} \times \lambda$ ($10^{-16}$ $\Omega$ m$^2$) | $\rho_{yy} \times \lambda$ ($10^{-16}$ $\Omega$ m$^2$) | $\rho_{zz} \times \lambda$ ($10^{-16}$ $\Omega$ m$^2$) | $\rho_{avg} \times \lambda$ ($10^{-16}$ $\Omega$ m$^2$) |
|---|---|---|---|---|---|---|---|---|---|
| BeCo | mp-2773 | Cubic | O [41, 42] | 0.376 | 4.42 | 4.76 | 4.76 | 4.76 | 4.76 |
| CoH | mp-1206874 | Cubic | X | 1.194 | 3.42 | 5.41 | 5.41 | 5.41 | 5.41 |
| CoPd$_3$ | mp-1183687 | Cubic | X | 3.086 | 4.30 | 6.49 | 6.49 | 6.49 | 6.49 |
| CoPt | mp-949 | Tetragonal | O [43-45] | 2.258 | 5.34 | 7.13 | 5.74 | 5.74 | 6.20 |
| CoPt$_3$ | mp-922 | Cubic | O [46, 47] | 2.886 | 5.60 | 5.51 | 5.51 | 5.51 | 5.51 |
| Co$_2$Tc$_6$ | mp-865733 | Hexagonal | X | 0.202 | 5.89 | 8.37 | 8.38 | 4.69 | 7.15 |
| Co$_3$Ni | mp-1008349 | Cubic | X | 5.656 | 4.68 | 5.79 | 5.78 | 5.78 | 5.78 |
| Co$_4$H | mp-1226072 | Trigonal | X | 6.048 | 4.17 | 6.62 | 6.62 | 6.69 | 6.64 |
| Co$_6$H$_2$ | mp-1025425 | Hexagonal | X | 8.76 | 4.28 | 6.55 | 6.55 | 9.66 | 7.59 |
| FeCo | mp-2090 | Cubic | O [48, 49] | 4.554 | 4.89 | 7.69 | 7.69 | 7.69 | 7.69 |
| MnCo | mp-1009133 | Cubic | X | 4.025 | 4.42 | 8.32 | 8.32 | 4.20 | 6.95 |
| Mn$_6$Co$_2$ | mp-1185970 | Hexagonal | X | 0.134 | 4.13 | 7.90 | 7.90 | 7.90 | 7.90 |
| Nb$_2$Co$_4$ | mp-670 | Cubic | O [50-52] | 1.435 | 5.30 | 4.76 | 4.76 | 4.76 | 4.76 |

[a] The synthesizability of the materials was determined by referencing the compositions provided in the ICSD.
[b] The cohesive energies are obtained from the Materials Project database.

a systematic approach was employed based on crystal symmetry. In the case of materials with cubic symmetry, the diagonal tensor components are identical, and this single value was utilized directly. For all non-cubic structures, an effective $\rho_{avg} \times \lambda$ value was calculated by arithmetically averaging the two in-plane and one out-of-plane tensor components to provide a measure of transport performance. Furthermore, compounds with $\rho_{avg} \times \lambda$ exceeding $15 \times 10^{-16}$ $\Omega m^2$ were removed from the dataset, as their electrical performance was considered insufficient for advanced interconnect applications.

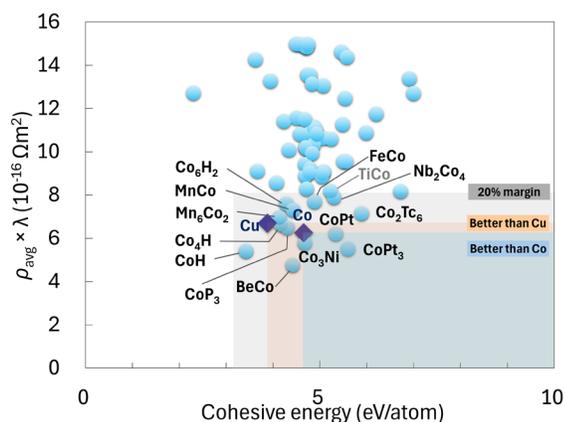

**Fig. 2** Performance distribution of Co compounds. Industrial reference material Cu and reference material Co are marked with purple diamonds. The shaded areas represent parameter space where performance is expected to be superior to each reference material. Note that compounds with $\rho_{avg} \times \lambda$ over $15 \times 10^{-16}$ $\Omega m^2$ are omitted due to their predicted poor electrical performance.

The results of this comprehensive screening are summarized in Fig. 2. The shaded regions establish performance benchmarks based on elemental Cu (the industry standard) and Co (reference for this study). To ensure a more comprehensive identification of promising candidates, we selected materials within a grey-shaded region representing a 20% performance margin relative to the Cu benchmark. This margin was established to broaden the search space, recognizing that practical interconnect integration requires a balance of properties. Candidates falling within this range, even if exhibiting slightly inferior metrics compared to Cu, may possess superior secondary attributes—such as thermal stability, adhesion strength, or process compatibility—that are critical for reliable device fabrication. Compounds located within this 'better' region are predicted to be promising, marking them as ideal candidates for future interconnect technology. Note that several candidates are located within the 'Better than Cu' region. This suggests they possess significant potential as next-generation interconnect materials.

It is also worth discussing the case of CoTi. Though not selected as a primary candidate in our screening, its calculated properties, $\rho_{avg} \times \lambda = 8.18 \times 10^{-16}$ $\Omega m^2$ and $E_{coh} = 5.23$ eV, place it near our performance benchmarks. This observation is interesting because CoTi has already been experimentally investigated as a reliable barrier/liner, capable of forming a robust 3 nm-thick layer due to its strong adhesion to SiO$_2$ [38–40]. The proximity of its calculated resistivity to our criteria suggests that, in addition to its established barrier function, CoTi may also possess favorable electrical performance, suggesting its potential as an interconnect material.

A comprehensive list of the 13 most promising compounds and their calculated properties is presented in Table 1. Notably, as indicated in the table, several candidates–including BeCo [41, 42], CoPt [43–45], CoPt$_3$ [46, 47], FeCo [48, 49], and Nb$_2$Co$_4$ [50–52]–have been experimentally synthesized, as verified through experimental reports within the inorganic crystal structure database (ICSD). This established synthesizability provides a credible pathway for the experimental realization of these materials in next-generation interconnects. Especially, it is worth noting that CoPt has already been experimentally reported to offer better resistivity than Cu at thin dimensions ($\leq$ 10 nm) [53].

Although only a few compounds outperform elemental Co in both metrics simultaneously, several candidates exceed Cu, demonstrating clear potential for practical interconnect applications. The successful identification of multiple candidates surpassing the industry-standard Cu demonstrates the effectiveness of our high-throughput screening approach as an effective tool for discovering novel interconnect materials. Note that our selection included candidates within a margin relative to the Cu benchmark. Even if the bulk properties of certain candidates do not outperform those of Cu, they may offer advantages in highly scaled dimensions. Unlike elemental Cu, which necessitates high-resistivity barrier/liner layers, Co-based compounds with relatively high cohesive energies are less prone to diffusion into oxides or Si substrates [40, 54]. This characteristic allows for barrier-less integration or significantly reduced barrier/liner thickness. Consequently, at highly scaled dimensions where barrier layers consume a significant portion of the cross-sectional area, the effective resistance increase in these materials is expected to be smaller than that of Cu.

Nevertheless, cautious investigation is warranted for specific candidates regarding safety and feasibility. For instance, compounds containing beryllium (Be) pose toxicity risks, while those containing technetium (Tc) or plutonium (Pu) involve concerns related to radioactivity. Although these factors may limit their adoption in general consumer electronics, such materials could still be considered for specialized applications—such as aerospace systems or extreme environments—where human exposure is strictly controlled or rigorous encapsulation is feasible [55, 56]. Regarding candidates not yet reported experimentally, it is important to note that they passed our thermodynamic stability criterion ($E_{\text{Hull}} \leq 20$ meV/atom), suggesting a high probability of future synthesis. Furthermore, the prevalent ferromagnetic nature of the candidates necessitates experimental verification, as it could induce magnetic crosstalk in highly scaled interconnects.

## 4 Conclusion

In this work, we conducted a high-throughput screening of Co-based binary compounds to find alternative interconnect materials to overcome the scaling limitations of Cu. We identified 13 promising candidates from a down-selected pool of 143 materials. Although strictly only six candidates fall within the "Better than Cu" regime based on bulk metrics, we emphasize that while resistivity and cohesive energy are key performance indicators, they are not the exclusive determinants of material applicability. By applying a rational margin, we extended our selection to include candidates that may offer superior integration feasibility and reliability in realistic scaled structures. Consequently, these compounds are expected to exhibit suppressed resistivity scaling in narrow interconnects and enhanced reliability against electromigration and diffusion. This study expands the scope of interconnect research beyond elemental metals. The results demonstrate that high-throughput screening is a powerful methodology for investigating the vast space of chemical substances, and future work will extend this approach to explore an even wider range of binary and ternary systems to accelerate the discovery of next-generation materials.

**Author Contributions** Gyungho Maeng: Data curation, Formal analysis, Investigation, Methodology, Software, Validation, Visualization, Writing – original draft. Yeonghun Lee: Supervision, Conceptualization, Project administration, Formal analysis, Funding acquisition, Methodology, Validation, Writing – review & editing.

**Funding** This work was supported by the National Supercomputing Center with supercomputing resources including technical support (No. KSC-2024-CRE-0069).

**Data Availability** The data that support the findings of this study are available from the corresponding author upon reasonable request.

## Declarations

**Conflict of interest** The authors declare they have no financial interests.

Electronic Materials Letters
https://doi.org/

**Graphical Abstract**

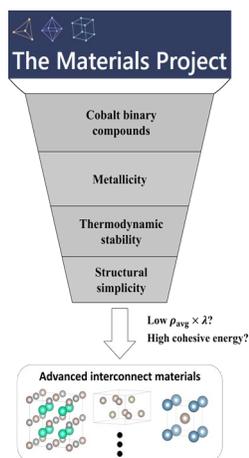